%% file: nayak_chitrachimera.tex
\documentclass[graybox]{svmult}

\usepackage{mathptmx}       
\usepackage{helvet}         
\usepackage{courier}        
\usepackage{type1cm}        
\usepackage{makeidx}         
\usepackage{graphicx}        
\usepackage{multicol}        
\usepackage[bottom]{footmisc}

\makeindex             

\begin{document}

\title*{Chimera states in  coupled  sine-circle map lattices}
\author{Chitra R Nayak and Neelima  Gupte}
\institute{Chitra R Nayak \at Department of Physics and Atmospheric Science, Dalhousie University, Halifax, NS, Canada. \\ \email{rchitra.r@gmail.com}
\and Neelima Gupte \at Department of Physics, Indian Institute of Technology, Madras, India. \\ \email{gupte@physics.iitm.ac.in}}
%
%
\maketitle

\abstract{Systems of coupled oscillators have been seen to exhibit chimera states, i.e. states  where the system splits into two groups where one group is  phase locked  and the other is  phase randomized. In this work, we report the existence of chimera states in a system of two interacting populations of  sine circle maps. This system also exhibits the clustered chimera behavior seen earlier in delay coupled systems. Rich  spatio-temporal behavior is seen in different regimes of the phase diagram. We carry out a detailed analysis of the stability regimes and map out the phase diagram using numerical and analytic techniques.
}

\section{Introduction}
\label{intro}
Coupled oscillator systems have long been studied as  good models
of a variety of experimental and natural systems, as well as paradigmatic systems for the observation and analysis of complex spatio-temporal behaviour. The study of synchronised behaviour in systems of identical
oscillators, as well as that of oscillators with frequencies drawn from a distribution, has resulted in important insights into the behaviour
of experimental systems, and has also contributed important techniques for
the analysis of extended dynamical systems.  
In addition to this, a  very interesting spatio-temporal pattern which is distinct from synchronised behaviour was reported by Kuramoto et al. in the case of identical oscillators with symmetrical coupling \cite{kuramoto:NPCS:02}. It was shown that under certain initial conditions and parameter values, an array of identical oscillators split into two groups: one showing coherent behavior and another incoherent behavior. Such states were named as chimera states by Strogatz et. al  \cite{abrams:prl:04}. The splitting of globally coupled oscillator population   with distributed frequencies into  synchronized and  desynchronized parts was earlier reported \cite{kuramoto:84,strogatz:pd:00,liang:chaos:10}. Various theoretical and numerical results have been reported since this discovery\cite{bordy:10,liang:10}.  An exact result for the stability and bifurcations of a system 
of two interacting populations of oscillators has been obtained \cite{abrams:prl:08}. It was shown that in a globally coupled network of oscillators with delay feedback, the single spatially connected region was replaced by a number of spatially disconnected regions of coherence with regions of incoherence in between. These states were named clustered chimera states\cite{sethia:prl:08,sheeba:pre:09}.   Chimera states have been described
as the natural link between coherent and incoherent states \cite{chenko:prl:08}. The spatiotemporal pattern for the coherent and incoherent regions forming the chimera states has been studied \cite{omel:10}.
\begin{figure}[t]
\sidecaption
\includegraphics[scale=.35]{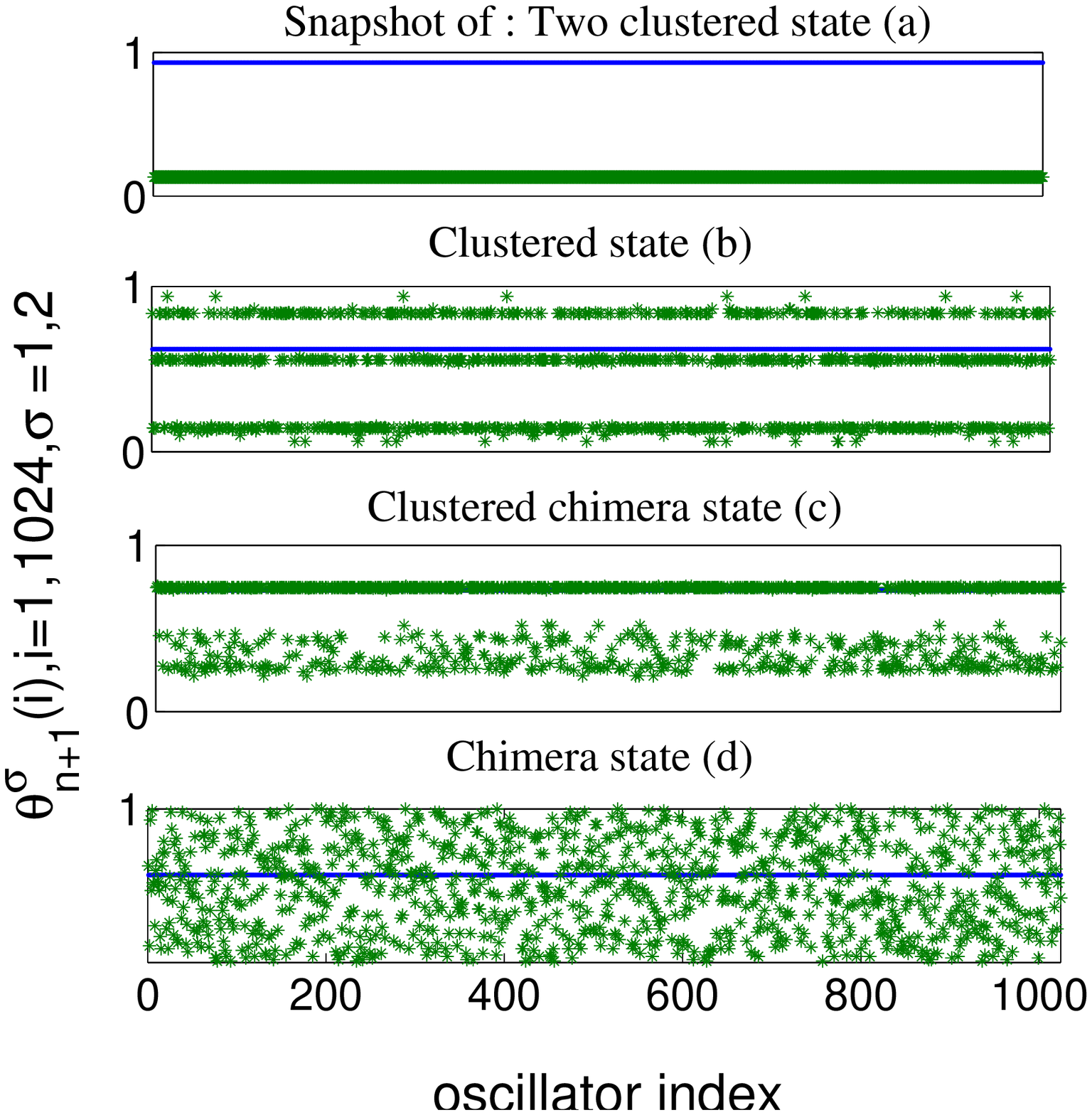}
\vspace{0.05cm}
\includegraphics[scale=.35]{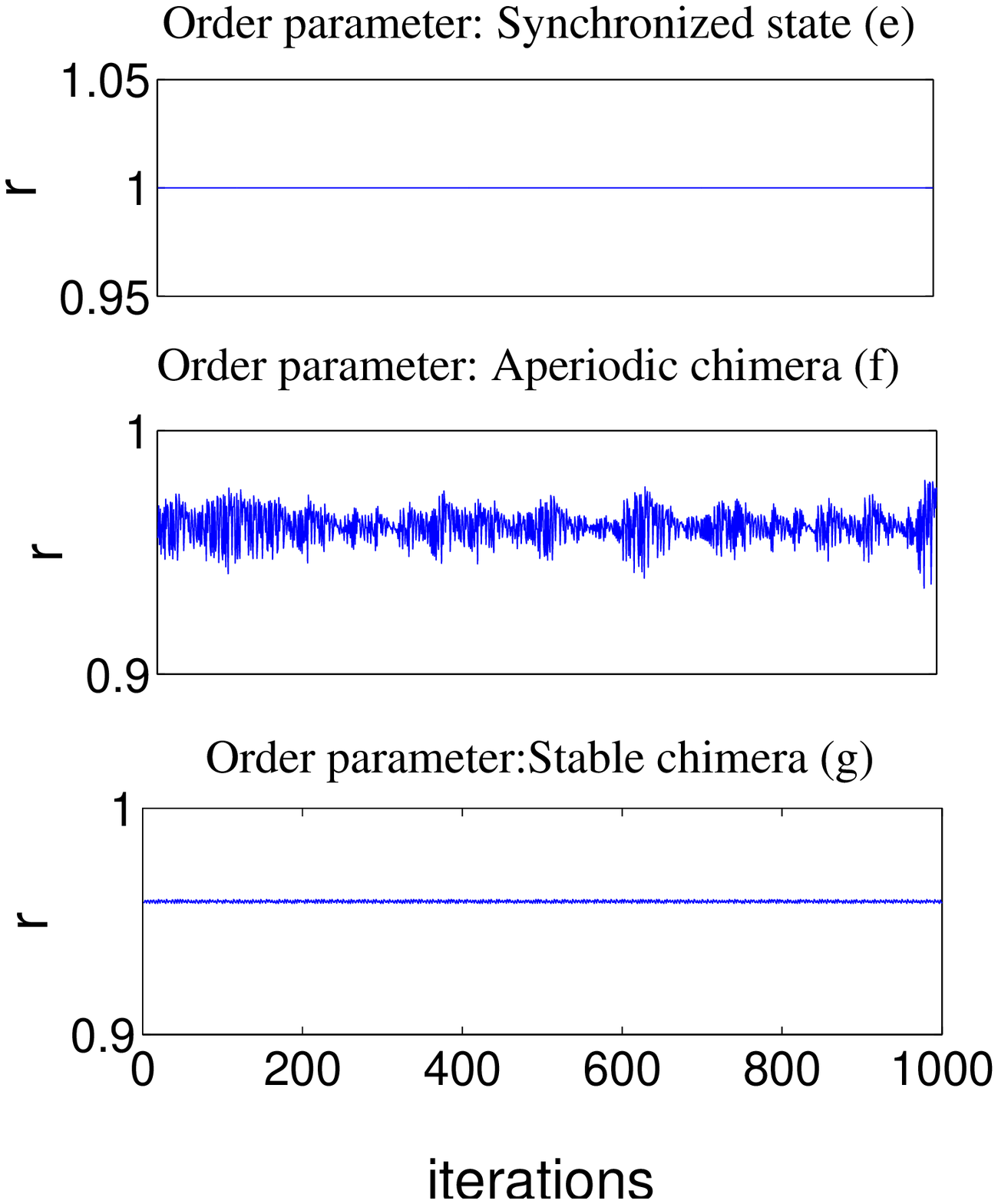}%
\caption{ Snapshots of the oscillators of the two populations in (a) two clustered state (b) clustered state (c) clustered chimera (d) chimera. The blue lines (thin line) correspond to group 1 where all the oscillators are synchronised. In (c) the oscillators from group 2 overlap with group 1. The order parameter for (e) the completely synchronized state (f) aperiodic chimera  (g) stable chimera. } %
\label{fig1}
\end{figure}
In this work, we study the existence and stability of chimera states  in two interacting populations of coupled sine circle maps. 
These systems are 
the map analogs of coupled oscillator systems. Coupled map systems show 
many of the varied phenomena observed in the case of continuous extended systems, but are far more numerically and analytically tractable. Coupled map systems have been seen to show synchronised solutions, spatiotemporal intermittency, spatiotemporal chaos \cite{jab:05,jan:03}. However, to the best of our knowledge,
chimera states have not been reported earlier in coupled map lattice systems.

\section{The model}

The single sine circle map evolves via the evolution equation
\begin{equation}
\theta_{t+1}=f(\theta_t)=\theta_t+\Omega-\frac{K}{(2 \pi)}\sin(2 \pi \theta_t)\phantom{aaa} mod 1
\end{equation}
Here,  $K$ is the nonlinearity parameter and $\Omega$ is the winding number of a single sine circle map in the absence of the nonlinearity.
This map shows a tendency to mode lock as the parameter $K$ is increased  and the phenomena of Arnold tongues are seen in the 
$K-\Omega$ space \cite{jensen:pra:84}. The winding numbers of the mode locked tongues show a
Devil's staircase structure.  There are several studies on  the spatiotemporal dynamics of diffusively coupled sine circle map lattices on regular sites \cite{nandini:pre:96}. 
The model that we consider  consists of two interacting populations of sine circle maps with every element of one population  coupled via  a coupling parameter $\epsilon_1$ to all the other elements in the same group and via another parameter $\epsilon_2$  to all those in the other group.  The evolution equation is given by
\begin{eqnarray} \nonumber
 \theta_{n+1}^{\sigma}(i)&=& (2-\epsilon_1-\epsilon_2)(\theta_{n}^{\sigma}(i)
+\Omega-\frac{K}{2 \pi}\sin(2 \pi \theta_{n}^{\sigma}(i))
+ \sum_{\sigma'=1}^{2}\frac{\epsilon_{\sigma \sigma'}}{N_{\sigma'}}\sum_{j=1}^{N_{\sigma'}}(\theta_{n}^{\sigma'}(j) \\
 &+&\Omega-\frac{K}{2 \pi}\sin(2 \pi \theta_{n}^{\sigma'}(j)) \phantom{aaa}mod 1
 \label{model}
\end{eqnarray}
where $\epsilon_{1 1}=\epsilon_{2 2}=\epsilon_1$ and $\epsilon_{1 2}=\epsilon_{2 1}=\epsilon_2$.  $\theta(t)$ is the angle at time $t$ and lies between $0$ and $1$.   The parameters $\Omega$ and $K$ are taken to be uniform at each site. The values of $\epsilon_1$ and $\epsilon_2$ should lie between $0$ and $1$ and  $\epsilon_1 + \epsilon_2 = 1$. We considered $1024$ oscillators in each group and  the results remain the same  with fewer  oscillators. The first population is given identical initial conditions while the second set is given random initial conditions between $0$ and $1$. The system exhibits rich spatial dynamics depending upon the value of $K$ and $\Omega$ for this initial condition. The most common solution for large values of $K$  are the completely synchronized state and the two clustered state, where all the oscillators of each group have the same value  of the phase but the phase values for each group differ from each other are shown in Fig. \ref{fig1}(a).  As the value of $K$ decreases the oscillators of group two bifurcate to the clustered state as shown in   Fig.\ref{fig1}(b). With further decrease in the value of $K$  all the oscillators of group two completely desynchronise, which correspond to the chimera state seen in Fig.\ref{fig1}(d).  One interesting solution that we observed was the clustered chimera state where the group two oscillators again separate into a completely synchronized state while the rest of the population remains desynchronized Fig.\ref{fig1}(c).  In Fig. \ref{fig1}(c) half of the population of the second group has the same state value as the first group and hence overlaps and the rest remains distributed over the phase space.  For all the values of $K$ and $\Omega$ the oscillators of group one remain completely synchronized.

\begin{figure}[t]
\sidecaption
\includegraphics[scale=.35]{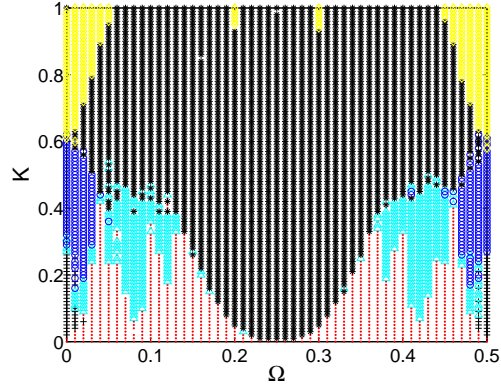}%
\caption{(Color online)The parameter space plot for $\epsilon_1 = 0.9$. The red $(\cdotp)$-s show chimera states, 
black $(+)$-s show clustered chimeras, cyan $( \bigtriangleup )$-s show clustered states,  blue $ (\circ) $-s show three 
clustered states, black $(\ast)$-s show two clustered states and yellow $(\diamond)$-s show complete synchronous states} %
\label{parachimepp9}
\end{figure}

The synchronization within the population of each kind can be characterised by the order parameter $r(t)=|\langle\exp(i \theta_j(t)) \rangle_2|$ where the angle brackets denotes the average over  the oscillators having the same dynamics. Thus for the completely synchronized, and the chimera state,  the average is taken over all the oscillators of group 2. The order parameter $r$ will be 1 for a completely synchronized state as the phases of all the oscillators are the same in this  state, and will take values between 0 and 1 when there is desynchronization or clustering. When $r$ takes a constant value the corresponding state is a stable state.  A time varying $r$ corresponds to  either  a breather solution or an unstable state. Chimera states can be statistically stable \cite{kuramoto:NPCS:02} in some cases leading to stable order parameters, in other cases the order parameter for chimera states can be oscillating or aperiodic  \cite{abrams:prl:08}. 
In our case, we had stable chimera states which bifurcate to the clustered state through an unstable aperiodic chimera state. Fig. \ref{fig1}(e) shows the order parameter for the group 2 oscillators for the completely synchronized state and Figs.\ref{fig1}(f) and (g) shows the aperiodic and stable chimera respectively.  For the clustered chimera and the clustered state  there is further division in the population and $r$ has to be defined separately for each group.

We explore the phase diagrams of the system at $K-\Omega-\epsilon$ values which exhibit the variety of possible solutions and also provide clues to the bifurcation sequences that take place.

\section{The phase diagrams and stability analysis}

\subsection{The $K-\Omega$ phase space}

We explore the $K-\Omega$ phase space with $\epsilon_1=0.9$, where a good spectrum of dynamical behaviour is seen  and map out the 
observed types of spatial behavior. The value of $K$ is varied from 0 to 1 and $\Omega$ from 0 to 0.5. The phase diagram is symmetric  about $\Omega=0.5$, hence values beyond $\Omega=0.5$ have not been plotted. Fig.\ref{parachimepp9} is the $K-\Omega$ parameter space plotted for $\epsilon_1=0.9$.  Low values of $K$  favor chimera states whereas for high values of $K$ the system exhibits  ordered spatial behavior.  For values of  $K$  near $1$ 
the system either is either in a  completely synchronized state or a two clustered state depending upon the $\Omega$ values for the initial conditions mentioned before. The yellow $(\diamond)$-s  and the black $(\ast)$-s represent the completely synchronised and two clustered state respectively and form the most observed states. Thus for a majority of parameter values the system is spatially ordered and forms a symmetric  pattern about $\Omega=0.25$. As the value of $K$ is decreased for values of $\Omega$ near $0$ and $0.5$ the oscillators of group 2 bifurcate into a three clustered state $\rightarrow$ clustered chimera $\rightarrow $ chimera state. In the remaining regions the clustered state bifurcates to a chimera state. Here the order parameter of the chimera state is aperiodic near the clustered state and then forms a stable chimera. The temporal behavior throughout the $\Omega-K$ space is chaotic except for the completely synchronized region.

Now we study three  $\epsilon-\Omega$ sections at three distinct values of $K$. One is the shift map case where $K=0$, another is the $K=1$ plane which is known to mode lock and the third is  a low value ( $K=0.1$), where chimera states are seen  and there is  rich  dynamical behavior.

\subsection{The Shift Map case}
The simplest case of sine circle maps is the shift map case with $K=0$.  For a single shift map given by
\begin{equation}
 \theta_{t+1}=\theta_t+\Omega \phantom{aa} mod\phantom{} 1
\end{equation}
the system has periodic orbits for rational values of $\Omega.$ The corresponding equation for the coupled system is 
\begin{eqnarray}
\theta_{n+1}^{\sigma}(i)=(\theta_{n}^{\sigma}(i)+\Omega)+ 
\sum_{\sigma'=1}^{2}\frac{\epsilon_{\sigma \sigma'}}{N_{\sigma'}}\sum_{j=1}^{N_{\sigma'}}(\theta_{n}^{\sigma'}(j)+\Omega )\hspace{0.3cm} mod 1
\end{eqnarray}
For the shift map case, the system always evolves to a chimera state for the initial conditions mentioned earlier. The numerically evaluated largest lyapunov exponent shows that the system is chaotic in the entire $\Omega-\epsilon_1$ phase space. The order parameter defined by $r(t)=|\langle \exp(i\theta_j(t))\rangle_2|$ remained constant with very small fluctuations which were due to the finite size effects. We checked this by varying the number of nodes in the constituting groups. Thus the shift map case has spatially stable chimera states that are temporally chaotic. 
The Jacobian matrix for the coupled system may be written as 

\[
J =
\left( \begin{array}{cc}
A & B \\
C & D \\
\end{array} \right) 
\]
 where 

 \[
A =
\left( \begin{array}{cccccc}
(2-\frac{N-1}{N} \epsilon_1 -\epsilon_2) f'(\theta_n^1(1)) & \frac{\epsilon_1}{N}f'(\theta_n^1(2))  & . & . & . & \frac{\epsilon_1}{N}f'(\theta_n^1(N))\\
\frac{\epsilon_1}{N}f'(\theta_n^1(1)) & (2-\frac{N-1}{N} \epsilon_1 -\epsilon_2) f'(\theta_n^1(2)) & . & . & . & \frac{\epsilon_1}{N}f'(\theta_n^1(N) \\
. & . & . & . & . &. \\
\frac{\epsilon_1}{N}f'(\theta_n^1(N) & \frac{\epsilon_1}{N}f'(\theta_n^1(N)) & . & . & . & (2-\frac{N-1}{N} \epsilon_1 -\epsilon_2) f'(\theta_n^1(2)) \\
\end{array} \right),
\]

 \[
D=
\left[ \begin{array}{cccccc}
(2-\frac{N-1}{N} \epsilon_1 -\epsilon_2) g'(\theta_n^2(1)) & \frac{\epsilon_1}{N}g'(\theta_n^2(2))  & . & . & . & \frac{\epsilon_1}{N}g'(\theta_n^2(N))\\
\frac{\epsilon_1}{N}g'(\theta_n^2(1)) & (2-\frac{N-1}{N} \epsilon_1 -\epsilon_2) g'(\theta_n^2(2)) & . & . & . & \frac{\epsilon_1}{N}g'(\theta_n^2(N) \\
. & . & . & . & . & .	 \\
\frac{\epsilon_1}{N}g'(\theta_n^2(N) & \frac{\epsilon_1}{N}g'(\theta_n^2(N)) & . & . & . &(2-\frac{N-1}{N} \epsilon_1 -\epsilon_2) g'(\theta_n^2(2))  \\
\end{array} \right]. 
\]

\[
B =
\left( \begin{array}{cccccc}
\frac{\epsilon_2}{N}g'(\theta_n^2(1)) & \frac{\epsilon_2}{N}g'(\theta_n^2(2))  & . & . & . & \frac{\epsilon_2}{N}g'(\theta_n^2(N))\\
\frac{\epsilon_2}{N}g'(\theta_n^2(1)) & \frac{\epsilon_2}{N}g'(\theta_n^2(2))  & . & . & . & \frac{\epsilon_2}{N}g'(\theta_n^2(N))\\
. & . & . & . & . & . \\
\frac{\epsilon_2}{N}g'(\theta_n^2(1)) & \frac{\epsilon_2}{N}g'(\theta_n^2(2))  & . & . & . & \frac{\epsilon_2}{N}g'(\theta_n^2(N))\\
\end{array} \right), 
\]
\[
C =
\left( \begin{array}{cccccc}
\frac{\epsilon_2}{N}f'(\theta_n^1(1)) & \frac{\epsilon_2}{N}f'(\theta_n^1(2))  & . & . & . & \frac{\epsilon_2}{N}f'(\theta_n^1(N))\\
\frac{\epsilon_2}{N}f'(\theta_n^1(1)) & \frac{\epsilon_2}{N}f'(\theta_n^1(2))  & . & . & . & \frac{\epsilon_2}{N}f'(\theta_n^1(N))\\
. & . & . & . & . & . \\
\frac{\epsilon_2}{N}f'(\theta_n^1(1)) & \frac{\epsilon_2}{N}f'(\theta_n^1(2))  & . & . & . & \frac{\epsilon_2}{N}f'(\theta_n^1(N))\\
\end{array} \right), 
\]
We have $f'(\theta_n^1(N))=1-K \cos(2 \pi \theta^1_n(N))$ and $g'(\theta_n^2(N))=1-K \cos(2 \pi \theta^2_n(N))$. The eigen values need to be numerically obtained since the stable state is the chimera state.

\begin{figure}[t]
\sidecaption
\includegraphics[scale=.35]{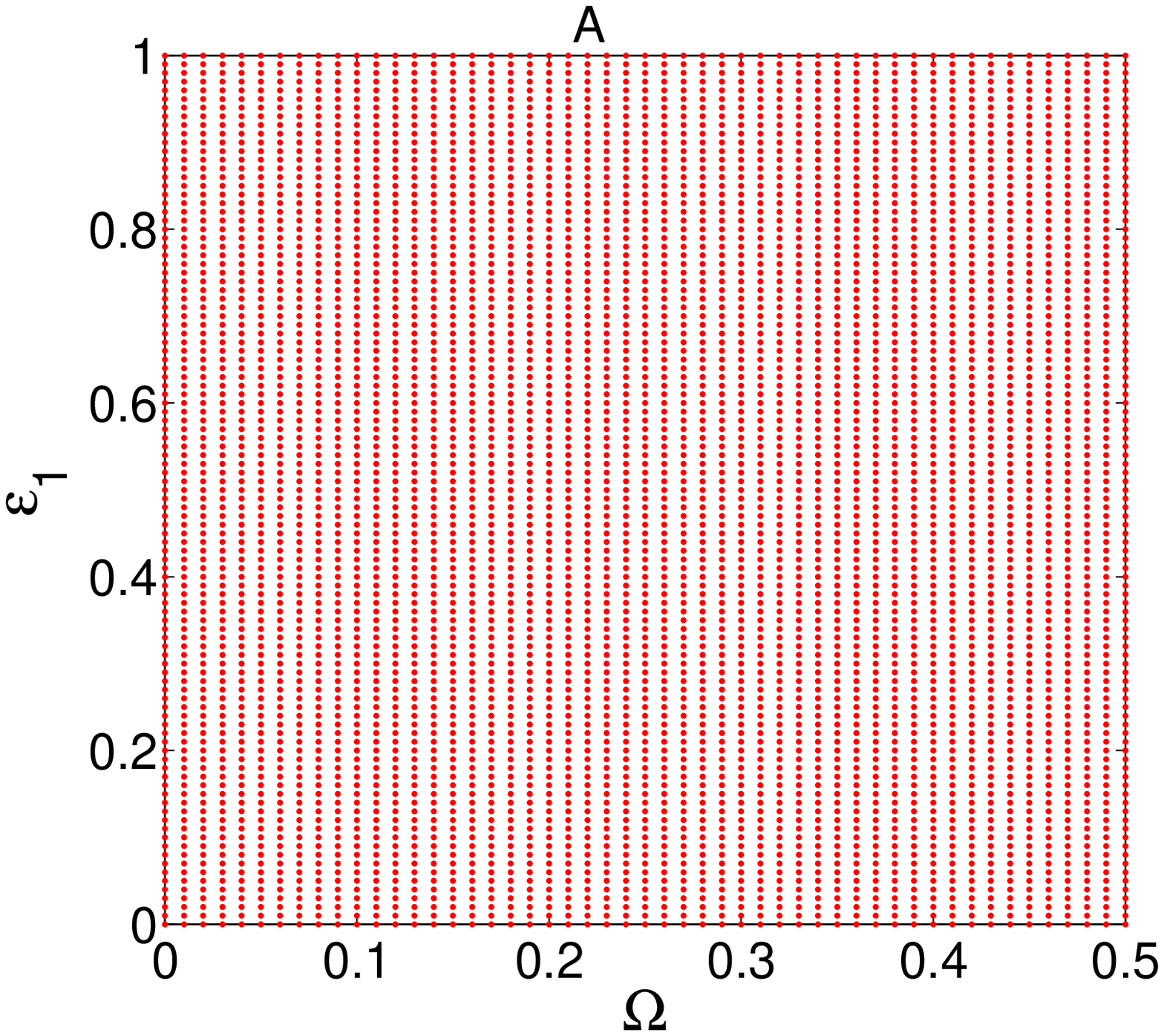}
\vspace{0.05cm}
\includegraphics[scale=.35]{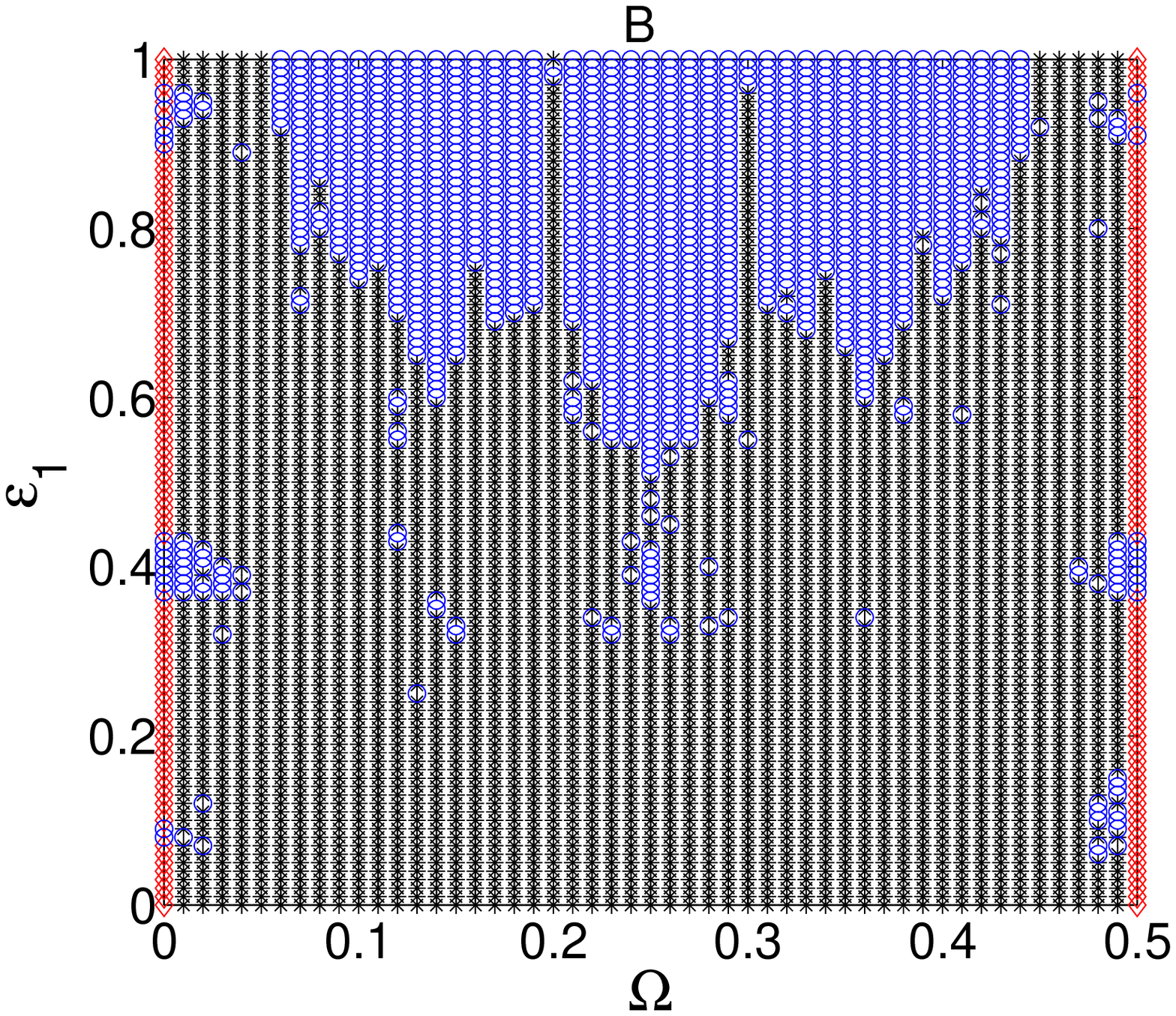}%
\caption{(Color online)A) The parameter space for the shift map case for the coupled sine-circle maps. Chimera states exists for all values of $\Omega \& \epsilon_1.$  B)The parameter space plot for $K=1$ . The red $(\bigtriangleup )$-s show completely synchronized fixed point solutions, 
black $(*)$-s show completely synchronized  solutions,   blue $ (\circ) $-s shows two clustered chaotic solutions. For $K=1,$ there are no chimeras.
} %
\label{k1p0}
\end{figure}

\begin{figure}[t]
\sidecaption
\includegraphics[scale=.35]{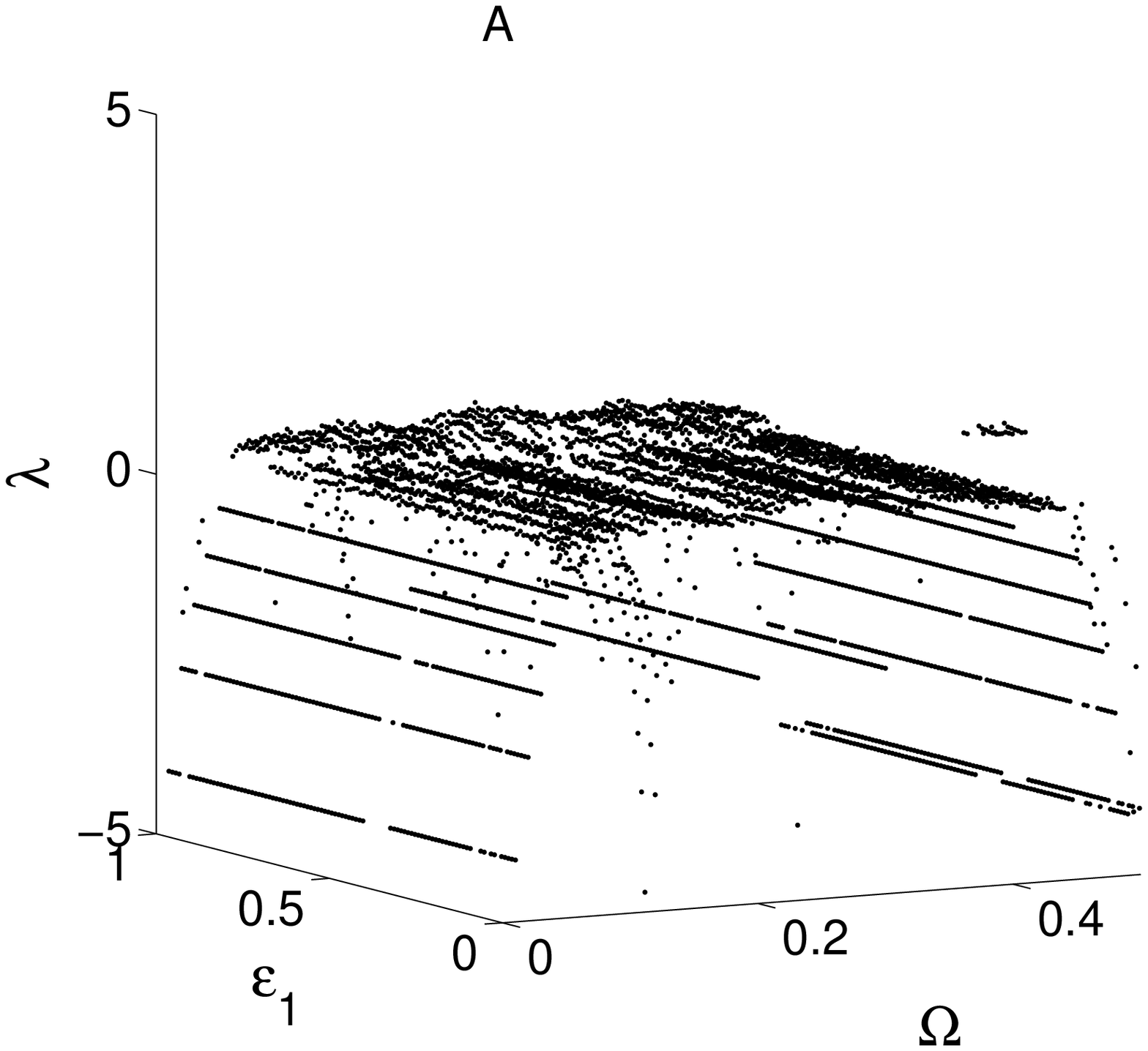}
\vspace{0.05cm}
\includegraphics[scale=.35]{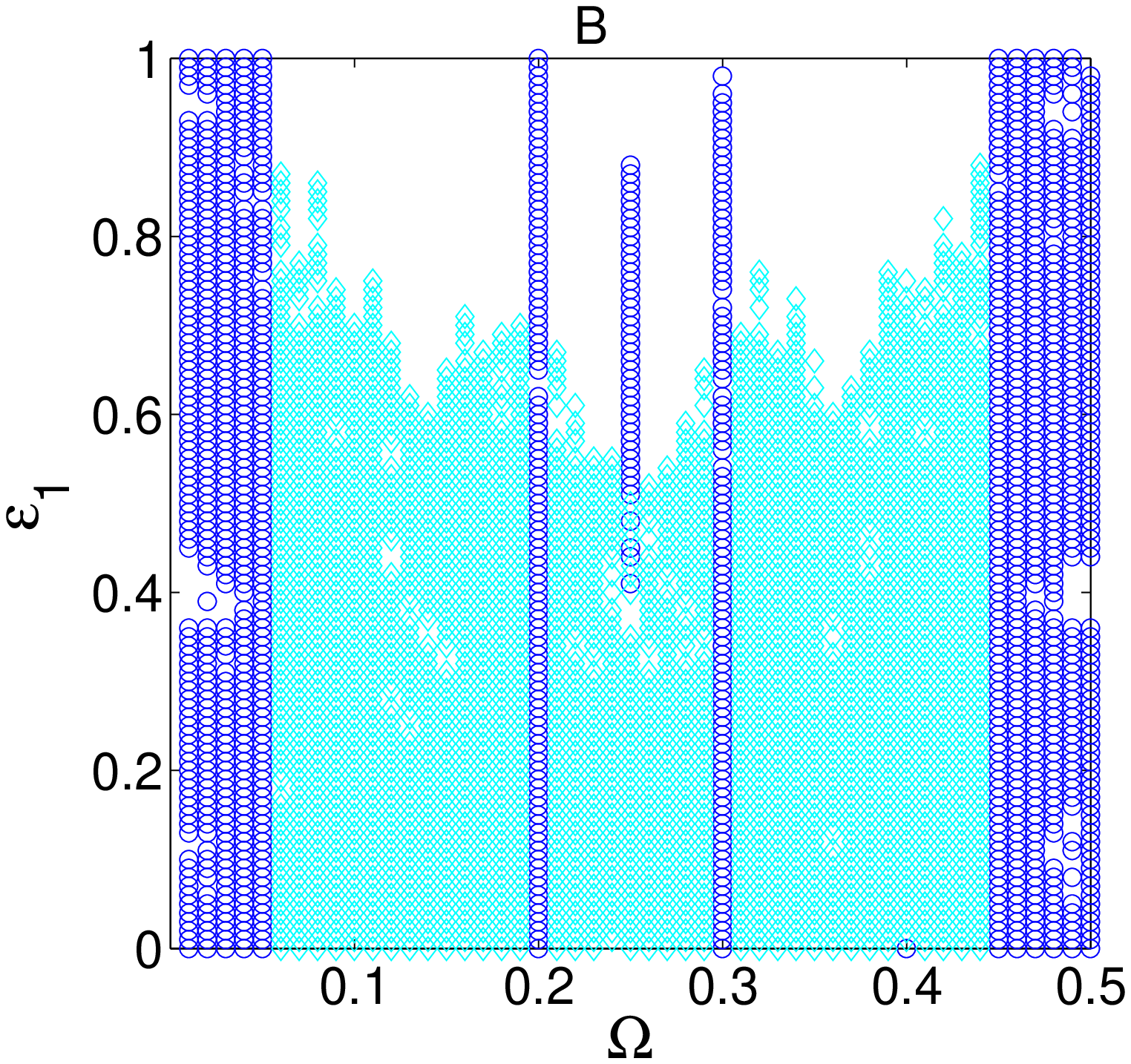}%
\caption{A ) The largest Lyapunov exponent obtained numerically for $K=1.0$. B) LE evaluated from Eq. \ref{largest} for $K=1.$ The blue $( \circ )$-s represent the periodic solutions and cyan $ ( \bigtriangleup )$-s are the chaotic solutions} %
\label{lyak1p0}
\end{figure}

\subsection{Linear Stability Analysis for $K=1$}
\label{lsa}
 Fig.\ref{k1p0} shows that in the $\epsilon_1-\Omega$ phase space plane with $K=1$, the system is always having either a complete synchronized state or a two clustered state. We study this plane analytically to obtain the bifurcation boundary.
In the completely synchronized state
$\theta_n^1(i)=\theta_n^2(i)=\theta_n^*$ where $i$ varies from $1$ to $N$. In this case the Jacobian matrix reduces to  
\[
J' =
f'(\theta_n^*)\left[ \begin{array}{cc}
A' & B' \\
B' & A' \\
\end{array} \right] 
\]   
with 

{\large \[
A' =
\left( \begin{array}{cccccc}
2-\frac{N-1}{N} \epsilon_1 -\epsilon_2  & \frac{\epsilon_1}{N}  & . & . & . & \frac{\epsilon_1}{N}\\
\frac{\epsilon_1}{N} & 2-\frac{N-1}{N} \epsilon_1 -\epsilon_2  & . & . & . & \frac{\epsilon_1}{N} \\
. & . & . & . & . &. \\
\frac{\epsilon_1}{N} & \frac{\epsilon_1}{N} & . & . & . & 2-\frac{N-1}{N} \epsilon_1 -\epsilon_2  \\
\end{array} \right),
\] }

and 

{\large \[
B' =
\left( \begin{array}{cccccc}
\frac{\epsilon_2}{N} & \frac{\epsilon_2}{N}  & . & . & . & \frac{\epsilon_2}{N}\\
\frac{\epsilon_2}{N} & \frac{\epsilon_2}{N}  & . & . & . & \frac{\epsilon_2}{N}\\
. & . & . & . & . & . \\
\frac{\epsilon_2}{N} & \frac{\epsilon_2}{N}  & . & . & . & \frac{\epsilon_2}{N}\\
\end{array} \right), 
\]}

This can be reduced   by a similarity  transformation to the form

\[
J' =
f'(\theta_n^*)\left[ \begin{array}{cc}
A'+ B' & 0 \\
0 & A'- B' \\
\end{array} \right] 
\] 
Now the Jacobian matrix has a block diagonalized form with each nonzero block being a circulant matrix such that

{\large \[
A'+B' =
\left( \begin{array}{ccccc}
2-\frac{N-1}{N} (\epsilon_1 +\epsilon_2))  & \frac{\epsilon_1+\epsilon_2}{N}  & . & . & \frac{\epsilon_1+\epsilon_2}{N}\\
\frac{\epsilon_1+\epsilon_2}{N} & 2-\frac{N-1}{N}( \epsilon_1 +\epsilon_2)  & . & .  & \frac{\epsilon_1+\epsilon_2}{N} \\
. & . & . & . & .  \\
\frac{\epsilon_1+\epsilon_2}{N} & \frac{\epsilon_1+\epsilon_2}{N}  & . & .  & 2-\frac{N-1}{N} (\epsilon_1 +\epsilon_2) \\
\end{array} \right),
\] }
and 
{\large \[
A'-B' =
\left( \begin{array}{ccccc}
2-\frac{N-1}{N} \epsilon_1 - \frac{N+1}{N}\epsilon_2  & \frac{\epsilon_1-\epsilon_2}{N}  & . & .  & \frac{\epsilon_1-\epsilon_2}{N}\\
\frac{\epsilon_1-\epsilon_2}{N} &  2-\frac{N-1}{N} \epsilon_1 - \frac{N+1}{N}\epsilon_2 & . & .  & \frac{\epsilon_1-\epsilon_2}{N}\\
. & . & . & . & .  \\
\frac{\epsilon_1-\epsilon_2}{N} 2-\frac{N-1}{N} & \frac{\epsilon_1-\epsilon_2}{N}  & . & .  & 2-\frac{N-1}{N} \epsilon_1 - \frac{N+1}{N}\epsilon_2\\
\end{array} \right),
\] }

This matrix has  eigenvalues  $2 f'(\theta_n^*) $,  $2 \epsilon_1 f'(\theta_n^*) $ and $(N-2)-$ fold degenerate eigen values $f'(\theta_n^*)$.
The Lyapunov exponents in terms of the eigen-values of the Jacobian matrix can  be written as 

\begin{equation}
\lambda_1=\frac{1}{\tau}\lim_{\tau\rightarrow\infty}\sum_{t=1}^{\tau}\ln\mid 2 f'(\theta_n^*)\mid
\label{largest}
\end{equation}

\begin{equation}
\lambda_2=\frac{1}{\tau}\lim_{\tau\rightarrow\infty}\sum_{t=1}^{\tau}\ln\mid 2 \epsilon_1 f'(\theta_n^*)\mid
\label{slargest}
\end{equation}
and
 \begin{equation}
\lambda_3=\lambda_4= ...= \lambda_N=\frac{1}{\tau}\lim_{\tau\rightarrow\infty}\sum_{t=1}^{\tau}\ln\mid  f'(\theta_n^*)\mid
\label{transverse}
\end{equation}

The numerically obtained value of the largest lyapunov exponent for the synchronised state agrees with 
the value given by Eq.\ref{largest}. Fig.\ref{lyak1p0}(a)
shows the numerical LE for the entire $\Omega-\epsilon_1$ and the system is chaotic for most of the values.  In the periodic regions, for $\Omega=0$ and $0.5$ the system has fixed point solutions and is periodic otherwise as seen 
in the figure.

\subsection{The $\Omega-\epsilon_1$ parameter space for $K=0.1$}

\begin{figure}[t]
\sidecaption
\includegraphics[scale=.35]{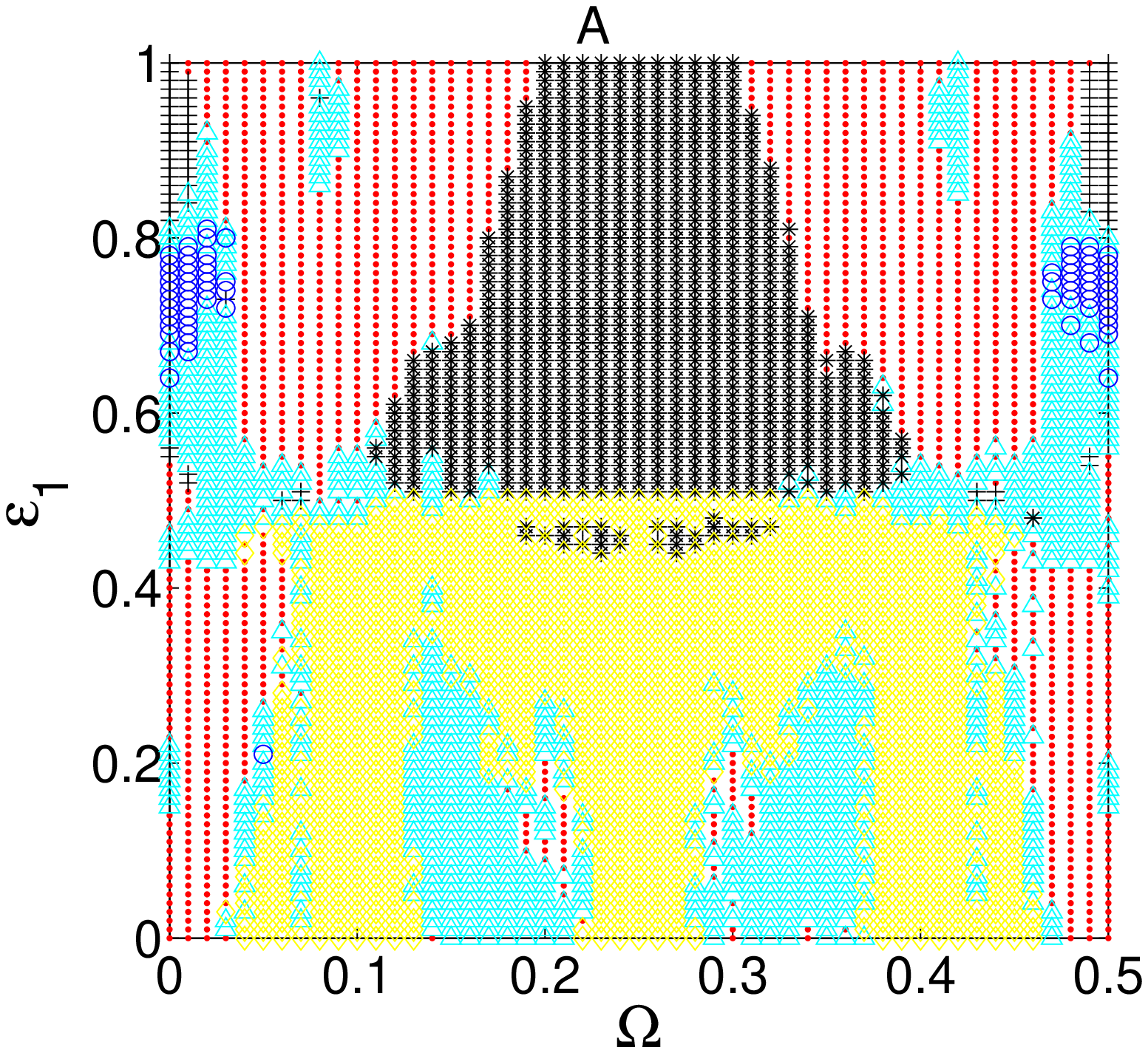}%
\vspace{0.05cm}
\includegraphics[scale=.35]{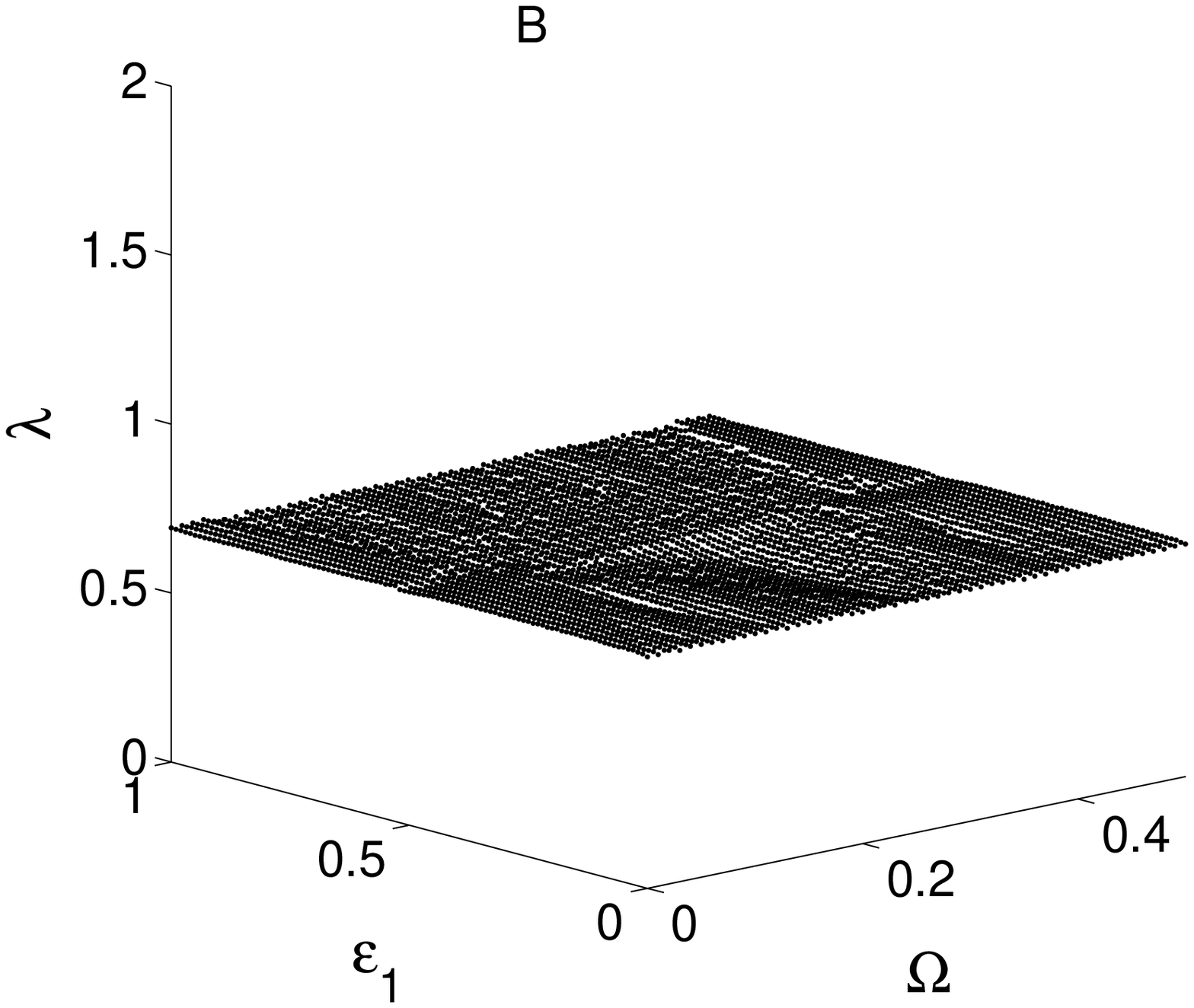}
\caption{(Color online) A) $\Omega-\epsilon_1$ plot with $K=0.1$. Red $( \cdot )$-s are the chimeras and black $ ( + )$-s are the clustered chimeras.The cyan $ ( \bigtriangleup )$-s shows clustered states, blue $( \circ ) $-s shows three clustered states, black $( \ast )$-s shows two clustered states and yellow $ ( \bigtriangleup ) $-s  shows complete synchronous states. B) The largest Lyapunov exponent shows that the system is always chaotic.
} %
\label{k0p1}
\end{figure}

The third slice along $K=0.1$  exhibits  rich spatial dynamics.
As noted earlier, chimera states are easily found at the lower values of $K$. For  $\epsilon_1$ above $0.5$ and for $\Omega$ near $0 \& 0.5$ the system has clustered chimera states (black $+$-s) which bifurcate to  clustered states (cyan $ \bigtriangleup $) as shown in Fig.\ref{k0p1} . For fairly large regions in the parameter space the system exhibits chimeras (red $\cdot$). The order parameter corresponding to the chimeras are by and large constants and hence stable chimeras are seen in this regime. The system also has two clustered solutions in this region which bifurcate to the unstable chimeras and then to stable chimeras. For values of $\epsilon_1$ less than $0.5$ in addition to these solutions the system also has a completely synchronized region (yellow $\bigtriangleup$). Thus, we note
that increasing the nonlinearity parameter $K$ tends to reduce the region
which is available for the chimera solutions. A $3-d$ phase diagram will 
illustrate this aspect further. We plan to explore this in future work.

\section{Conclusions}
In this paper, we proposed a couple map model with two interacting species of sine circle maps coupling. Chimera states are easily seen in this system 
given two sets of initial conditions for the two species, one random and one synchronised. Special choices of initial conditions are not required.  
The $K=0$ i.e. the shift map version of this models shows temporally chaotic chimeras all over
the parameter space. The introduction  of $K$ stabilises non-chimera states ,  but chimera states are still easily seen at low values. 
 Clustered chimera states where the desynchronized population again splits into a synchronized and desynchronized population are seen here, without the introduction of delays. 
The stability analysis for synchronised states is carried out analytically
and numerically, so that bifurcations from the synchronised values can be obtained.  
 We observe that the stable chimera bifurcates to the clustered state through an aperiodic chimera. The full $3-$ dimensional phase diagram remains to be explored, hence other bifurcation sequences for the chimeras 
may be possible. We hope to explore these directions in future work. 

\begin{acknowledgement}
NG thanks CSIR for partial support in this work.
Travel support was provided by the Office of Naval Research Global (ONR Global) to NG and CRN :under the Visiting Scientist Program.
\end{acknowledgement}
%

\input{referenc}

\end{document}

%% file: referenc.tex
%
%